\documentclass[12pt]{iopart}
\expandafter\let\csname equation*\endcsname=\relax 
\expandafter\let\csname endequation*\endcsname=\relax 
\usepackage{amsmath,amssymb,amsthm} 
\usepackage{braket}
\usepackage{graphicx}
\usepackage{xcolor}
\usepackage{graphicx}
\usepackage{dcolumn}
\usepackage{bm}

\usepackage{caption}

\begin{document}

\title[]{An unusual phase transition in a non-Hermitian Su-Schrieffer-Heeger model}

\author{A Niveth$^{1}$, S Karthiga$^{2*}$ and M Senthilvelan$^{1}$}
\address{$^{1}$Department of Nonlinear Dynamics, School of Physics, Bharathidasan University, Tiruchirappalli - 620 024, 
	Tamil Nadu, India. \\  $^{2}$PG and Research Department of Physics, Seethalakshmi Ramaswami College (Autonomous), Affiliated to Bharathidasan University, Tiruchirappalli - 620 002, Tamil Nadu, India.}
\ead{karthigasuthagar@gmail.com}
\vspace{10pt}

\begin{abstract}
\par This article studies a non-Hermitian Su–Schrieffer–Heeger (SSH) model which has periodically staggered Hermitian and non-Hermitian dimers. The changes in topological phases of the considered chiral symmetric model with respect to the introduced non-Hermiticity are studied where we find that the system supports only complex eigenspectra for all values of $u \neq 0$ and it stabilizes only non-trivial insulating phase for higher loss-gain strength. Even if the system acts as a trivial insulator in the Hermitian limit, the increase in loss-gain strength induces phase transition to non-trivial insulating phase through a (gapless) semi-metallic phase.  Interesting phenomenon is observed in the case where Hermitian system acts as a non-trivial insulator.  In such a situation, the introduced non-Hermiticity neither leaves the non-trivial phase undisturbed nor induces switching to trivial phase.  Rather, it shows transition from non-trivial insulating phase to the same where it is mediated by the stabilization of (non-trivial) semi-metallic phase.  This unusual transition between the non-trivial insulating phases through non-trivial semi-metallic phase gives rise to a question regarding the topological states of the system under open boundary conditions.  So, we analyze the possibility of stable edge states in these two non-trivial insulating phases and check the characteristic difference between them.  In addition, we study the nature of topological states in the case of non-trivial gapless (semi-metallic) region.   
\end{abstract}

%
%
%
%
%

\section{Introduction}
\par In recent years, much attention has been paid to topological materials which support states that are robust against moderate levels of disorder and perturbation \cite{rev_1, rev_2, nat_2}. One such material that has been reported first in the literature is topological insulators.  In this, besides bulk of the sample shows insulating behavior, their edges (or surface) show current conduction so they have hybrid features of insulators and conductors \cite{rev_2, nat_2, Asboth_shortcourse, PRB_108}.   The robustness of these system towards environmental perturbations allows promising applications in preparing robust entangled states \cite{Robust.Entang}, fault-tolerant quantum computing \cite{Quant.Memory}, spintronics \cite{Robust.T.states} and electronics \cite{PRR 5}.  In the earlier years, classification of these non-trivial insulators from trivial ones was unclear, but researches reveal that under the lens of topology (or topological invariant), the band structures can reveal the trivial/non-trivial nature of the system.  This led to the exploration of topological states in the realm of Hermitian lattices \cite{rev_1, rev_2, nat_2, Asboth_shortcourse, PRB_108}. 
\par  The physical realization of non-Hermitian systems \cite{ nH_SSH, nH_tetra_SSH,nh_rev} and their unique features which could not be achieved in Hermitian systems (such as, power oscillations \cite{longhi,PRA94}, unidirectional invisibility \cite{PRA104}, unconventional lasers \cite{PRA99} and exceptional point encirclement \cite{commun.P7}) stimulated the researchers to extend its domain to include topological condensed matter systems.  Apart from the theoretical findings \cite{rev_1, rev_2, nat_2, Asboth_shortcourse, PRB_108, nH_tetra_SSH, antiPT, SSH_og, PRB103, PRR5, PTS_with_comp._boundary_pot.}, the experimental studies on these dissipative systems have also shown the existence of novel topological states and unique behaviors \cite{nh_exp1, nh_exp2, nh_exp3}.  One of such unique behaviors is the non-Hermitian skin effect (NHSE) observed in certain non-Hermitian systems where localization of maximum number modes at one of the edges (a directional hopping) is observed.  Due to the above, the bulk-boundary correspondence ( which relates the bulk topological invariant characterizing the Bloch bands of a periodic system and the occurrence of edge-modes), a key principle in the topological systems, itself differs in many of the non-Hermitian systems.   Secondly, the symmetry classes categorizing topological system get ramified in the presence of non-Hermiticity (where the Hermitian systems fall within 10-fold Altland-Zirnbauer (AZ) symmetry classes but the non-Hermitian ones have 38-fold symmetry classification \cite{sym}).  With complex energy bands, the definition of energy gap in these situations also becomes non-trivial where it can be a line gap or point gap in the complex plane and the bands are said to be separable or isolated or inseparable. These complex energy bands exhibit interesting form of degeneracy at the spectral singularity, namely at the exceptional point (EP), where eigenfunctions and eigenvalues coalesce. The change in the eigenvalues from complex to real induces phase transition and this spectral singularity leads to many intriguing phenomena including topological chirality, loss-induced transparency and unidirectional invisibility \cite{ep1}. Particularly, higher order exceptional points are shown to enhance sensing capabilities of resonant optical structures \cite{ep1} and the real part of fidelity susceptibility density goes to negative infinity when the parameter approaches the EP \cite{ep2,ep3, Starkov}.

\par In \cite{prl_takata}, non-Hermiticity is shown to induce non-trivial insulating phase in a topologically trivial (gapless) system and to support edge modes and interface states.  Similarly, higher order topological insulating phase induced deliberately by non-Hermiticity is illustrated experimentally in acoustic crystals \cite{acous}.  In \cite{indian}, a comparison between the topological properties of $\cal{PT}$-symmetric case with the non-$\cal{PT}$-symmetric case is made, where the breakdown of bulk-boundary correspondence in the non-$\cal{PT}$-symmetric case is illustrated. An anti-$\cal{PT}$-symmetric model is studied in \cite{antiPT}, where the non-Hermiticity is shown to favor non-trivial topology.  The existence of novel topological states in lattices with multiple number of sites per unit cell \cite{nH_tetra_SSH,eight} and using different models like Su–Schrieffer–Heeger (SSH) model \cite{S.Lieu}, Aubry-Andr\'e model \cite{PRB103}, Hatano-Nelson model \cite{Ann.Rev.CMP} and Kiteav chain \cite{PRR5} are studied. In \cite{new11}, use of entanglement entropy in understanding the phase transitions in non-Hermitian systems is shown.

\par In this article, we consider SSH model as a toy model.  Instead of replacing all the sites of SSH lattice to be a non-Hermitian one \cite{nH_tetra_SSH, antiPT, CJP, PhysScri}, we have replaced the alternative unit cells of SSH with the non-Hermitian cells, so that, the particle hopping along the chain will experience alternatively Hermitian and non-Hermitian environments. With the reduced number of non-Hermitian sites, we study how far the non-Hermitian attributes are replicated in their topological states. The considered non-Hermitian model retains chiral symmetry of  SSH and it comes under the BDI$^\dagger$ symmetry class (which is unique to non-Hermitian systems).  Different topological phases of the system are studied with respect to non-Hermiticity. In the Hermitian limit, SSH supports trivial and non-trivial insulating phases in different parametric regions. But in the presence of strong non-Hermiticity, our system exhibits only non-trivial insulating phase in all the parametric regimes. We observe two different phase transitions, trivial to non-trivial insulating phase and non-trivial to non-trivial insulating phase which are mediated by (non-trivial) semi-metallic phase.  The gap closing through the emergence of exceptional points is responsible for the intermediate semi-metallic regime.  An interesting thing observed in this context is that gap closing is usually an indicator for switching of topological phases (trivial to non-trivial and vice versa) and semi-metallic phase is also shown to intermediate such topological switching \cite{semimetal}.  But here the semi-metal phase intermediates transition between two topologically equivalent phases. The non-Hermiticity here restores the non-trivial insulating phase.  Then there arises the question that whether these two topological states (that appear at the two extremes of gapless regime) are same in all perspectives or different.  We study the above in the situation with open boundary conditions and find the edge states in these topological regimes. 
\par To illustrate the above, the manuscript is structured in the following manner.   In Sec. \ref{sec_mod}, the model under consideration is given.   The different phases of the system under the periodic boundary conditions (PBC) are presented in Sec. \ref{sec_per}, where the band structure and corresponding topological invariant are shown with respect to different parameters.  Sec. \ref{sec_open} details the topological edge states in the open boundary condition (OBC) situation and presents the possibilities of NHSE.

\section{\label{sec_mod}Model}

\begin{figure}[ht!]  
	\captionsetup{justification=justified} 
	\includegraphics[width=1\linewidth]{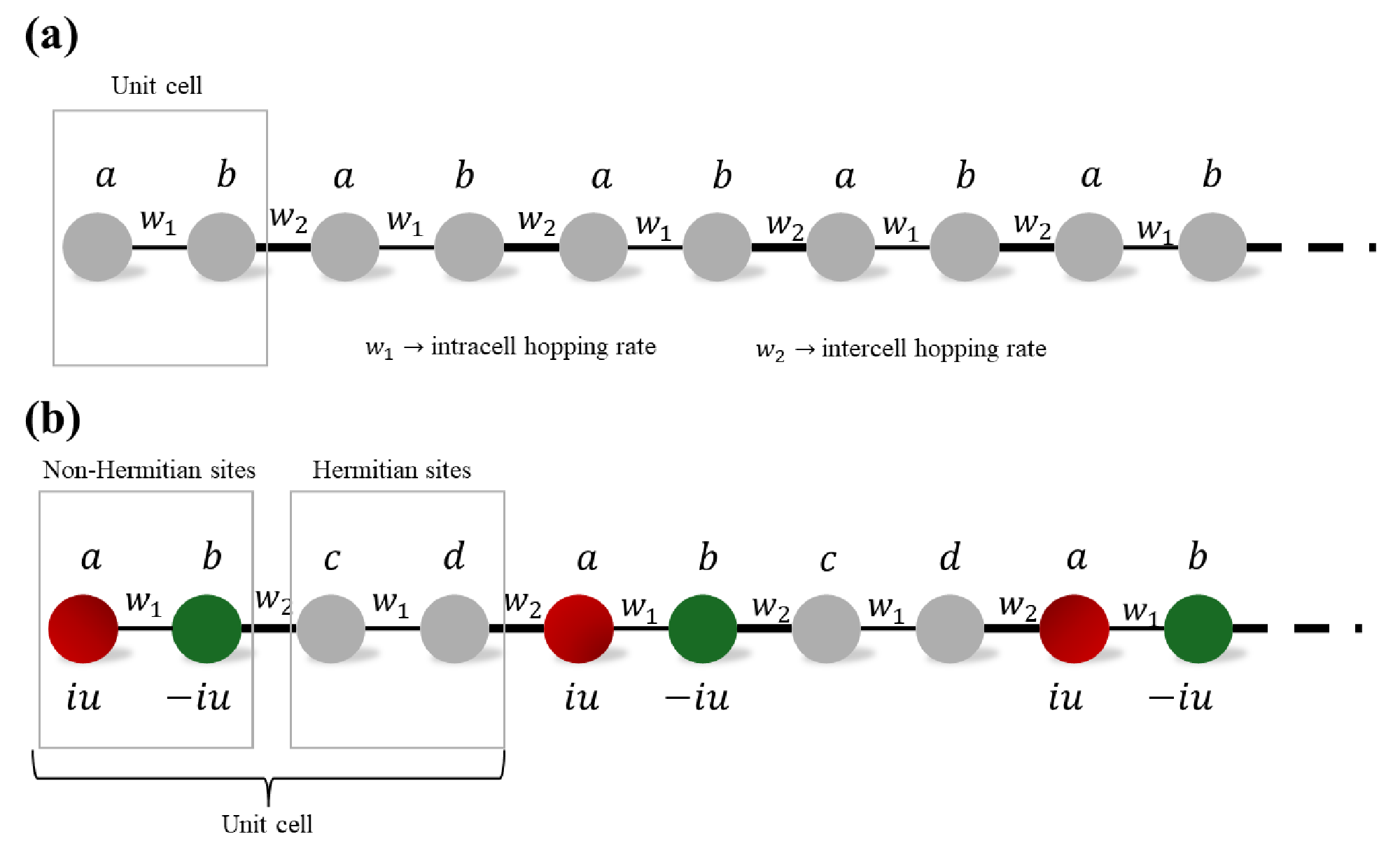}
	\caption{Schematic diagram: (a) SSH model, (b) non-Hermitian SSH model under consideration. Gray colored sites denote Hermitian sites and red and green colored ones represent non-Hermitian sites. }
	\label{model_fig}
\end{figure}

\par It is well known in the literature that the SSH model is the prototypical model for the investigation of topological phenomena in condensed matter physics \cite{Asboth_shortcourse}.  The SSH model (shown in Fig. \ref{model_fig}(a)) describing the motion of electron in polyacetylene chain has two sites per unit cell in which intercell hopping is different from the intracell hopping.  Hamiltonian of the above model can be given by,
\begin{equation}
		H_{\mathrm{SSH}} =  w_1 \left[\sum_{n=1} ^{N} (\ket{n,a}\bra{n,b}+\mathrm{h.c.})\right] + w_2 \left[\sum_{n=1} ^{N} (\ket{n,b}\bra{n+1,a}+\mathrm{h.c.})\right],
\end{equation}
where $w_1$ and $w_2$ are respectively intracell and intercell hopping integrals and are considered as $w_1=w(1-\delta\cos{\theta})$ and $w_2=w(1+\delta\cos{\theta})$ with the modulation intensity of hopping amplitudes $0<\delta<1$  and the phase parameter, $-\pi\leq \theta \leq \pi$. 
As mentioned in the introduction, the inclusion of non-Hermiticity in such model add feathers to the cap of the topological properties of SSH model \cite{nH_SSH,nH_tetra_SSH, antiPT, trimierized_optical}. Even though the considered SSH system has been first realized in the case of quantum systems, the similarities between the Schr\"odinger equation and the electromagnetic wave equation helped to realize the model in photonics. Considering the experimental feasibility, the theories of topological systems were tested using photonic systems including waveguide arrays and micro-ring resonators. Similarly, the non-Hermitian system which are studied in the case of open quantum problems which are modelled by quantum master equations \cite{SciPost,J.Opt.} can also be observed in the photonics. The non-Hermitian SSH models have no exclusion \cite{Exp} and they have been experimentally realized using the photonics platforms by suitably tailoring their loss and gain \cite{PhotLatt}.  Because, non-Hermiticity can be suitably introduced into the waveguide arrays through loss inducing chrome stripes or through the suitable radiative losses \cite{PhotLatt}.  The gain can be introduced in the waveguide arrays through the optical pumping of Fe-doped LiNbO$_3$ waveguides \cite{PhotLatt} and through the use of electrically pumped quantum dots.  Considering microrings, InGaAsP under optical pumping is used to induce gain and chrome layer is used to induce loss \cite{PhotLatt}. In this connection, we here consider a non-Hermitian SSH model in which the alternative unit cells of the standard Hermitian SSH model is replaced with non-Hermitian dimer, thus a periodic arrangement of Hermitian and non-Hermitian dimers can be seen in Fig. \ref{model_fig}(b). As mentioned above, the possibility of engineering loss and gain  in photonic lattices hints the experimental realization of the above system.
\par  Hamiltonian of the considered system can be written as 
\begin{equation}
	H = H_{hop} + H_{nh},
	\label{syst_eq}
\end{equation}
where, $H_{hop}$ defines the hopping among the sites (which is similar to SSH Hamiltonian) and $H_{nh}$ represents the onsite non-Hermitian potential. $H_{hop}$ can be written as
\begin{equation}
	\begin{split}
		H_{hop} = & w_1 \left[\sum_{n=1} ^{N} (\ket{n,a}\bra{n,b}+\ket{n,c}\bra{n,d}+\mathrm{h.c.}\right]\\& 
		+ w_2 \left[\sum_{n=1} ^{N} (\ket{n,b}\bra{n,c}+\ket{n,d}\bra{n+1,a}+\mathrm{h.c.}\right].
	\end{split}
\end{equation}
Here, $\ket{n,x}$ with $x\,=\,a,\,b,\,c$ and $d$, represent the state vectors of each sites in the $n^{\mathrm{th}}$ unit cell of the lattice and $N$ represents the total number of unit cells in the lattice. Onsite complex potentials of non-Hermitian sites can be given by,
\begin{equation}
	\centering
	H_{nh} = i u \left[ \sum_{n}^{N} [\ket{n,a}\bra{n,a} - \ket{n,b}\bra{n,b}] \right],
\end{equation}
where $u$ represents the loss-gain strength.  It is clear from the above that the sites $a$ and $b$ are non-Hermitian sites and $c$ and $d$ are Hermitian sites. For simplicity, we consider $w=1$ and $\delta=0.3$ for all the studies carried out in this article.

\section{\label{sec_per}Under Periodic Boundary Condition}
Under PBC, the bulk momentum space Hamiltonian of the system (\ref{syst_eq}) can be obtained through Fourier transformation as \cite{Asboth_shortcourse} 
\begin{equation}
	H_k(k) = 
	\begin{pmatrix}
		i u  & w_1 & 0 & w_2 e^{-ik}\\
		w_1 & -i u & w_2 & 0 \\
		0 & w_2 & 0 & w_1 \\
		w_2 e^{ik} & 0 & w_1 & 0\,
	\end{pmatrix},
	\label{H_matrix}
\end{equation}
where $k$ is the Bloch wave number. 

\par As topological properties of a system are entangled with their symmetries, we examine the invariance of the system with respect to different symmetry operations. We find that the system exhibits the non-Hermitian variant of time reversal symmetry, ($\mathrm{TRS}^\dagger$ symmetry),
\begin{equation}
	T_- H_k^* T_-^{-1} = - H_{-k}, \quad \mathrm{where}, \quad T_- = \kappa(\tau_z \otimes \sigma_0),
\end{equation}
$\kappa$ is complex conjugation operator and the non-Hermitian variant of Particle Hole symmetry ($\mathrm{PHS}^\dagger$ symmetry),
\begin{equation}
	C_+ H_k^T C_+^{-1} = H_{-k}, \quad \mathrm{where}, \quad C_+ = \tau_0 \otimes \sigma_0.
\end{equation} 
The presence of these two symmetries implies the chiral symmetric (CS) nature of the system,
\begin{equation}
	\Gamma H_k^\dagger \Gamma^{-1} = -H_k, \quad \mathrm{with} \quad \Gamma=\tau_z \otimes \sigma_0.
\end{equation}
In the above, $\sigma_i$ and $\tau_i$, $i=0$, $x$, $y$, and $z$, are respectively the Pauli matrices. From \cite{sym}, we can find that the model comes under $\mathrm{BDI}^{\dagger}$ class of symmetries. It may be noted that the systems of $\mathrm{BDI}^{\dagger}$ class of symmetries will show phase transition from trivial (non-trivial) to non-trivial (trivial) topological phases by band gap closing in the real part of energy \cite{sym, nH_tetra_SSH}.

\subsection{Band Structure}

\par Before studying the band structure and topological phases of the considered model, let us recall those in the case of classical SSH model \cite{Asboth_shortcourse}, where one can find that the bands are gaped in two cases (i) the case in which the intercell coupling is stronger, that is, $w_2>w_1$ or $-\frac{\pi}{2} \leq \theta \leq \frac{\pi}{2}$ and (ii) the case in which the intracell coupling is stronger, ie., $w_2<w_1$ or $\frac{\pi}{2} \leq |\theta|< \pi$ .  Band merging is observed at $\theta=\frac{\pi}{2}$.  The insulating phases observed in the above mentioned two cases are not topologically equivalent.  The Zak phase ($\Omega$) calculated for different values of $\theta$ clearly shows that the non-trivial values of $\Omega$ for $-\frac{\pi}{2} \leq \theta \leq \frac{\pi}{2}$ indicates non-trivial phase and trivial values in the region $\frac{\pi}{2} \leq |\theta| \leq \pi$ indicates trivial phase \cite{nH_tetra_SSH}.

\par
\begin{figure}[h!]
	\captionsetup{justification=justified}
	\begin{center}
		\includegraphics[width=1.0\linewidth]{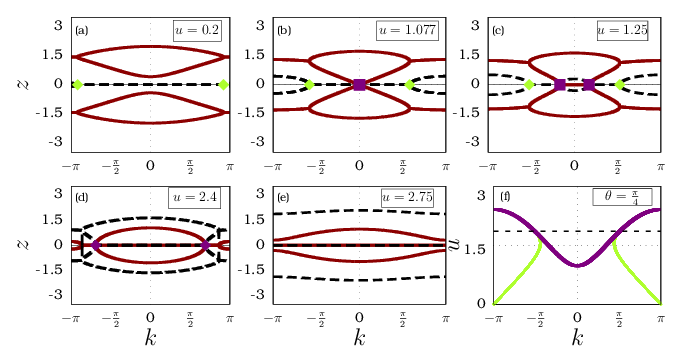}
		\caption{(a)-(e) show the band structures corresponding to the case $\theta=\frac{\pi}{4}$ for different value of $u$, namely, $u=0.2$, $1.077$, $1.25$, $2.4$, and $2.75$.   The continuous dark red curve represents the $\mathrm{Re}[z(k)]$ and the dashed black curve represents $\mathrm{Im}[z(k)]$. (f) shows the trajectories of exceptional points $\mathrm{EP}_1$ (shown by purple curve) and $\mathrm{EP}_2$ (shown by yellowish green curve)  and the dashed line is drawn for $u=u_m$ where $\mathrm{EP}_1$ and $\mathrm{EP}_2$ coalesce.  The yellowish green colored diamond and the purple colored dot respectively represent the exceptional points $\mathrm{EP}_1$ and $\mathrm{EP}_2$. }
		\label{band_11}
	\end{center}
\end{figure}

\begin{figure}[h!]
	\captionsetup{justification=justified}
	\begin{center}
		\includegraphics[width=1.0\linewidth]{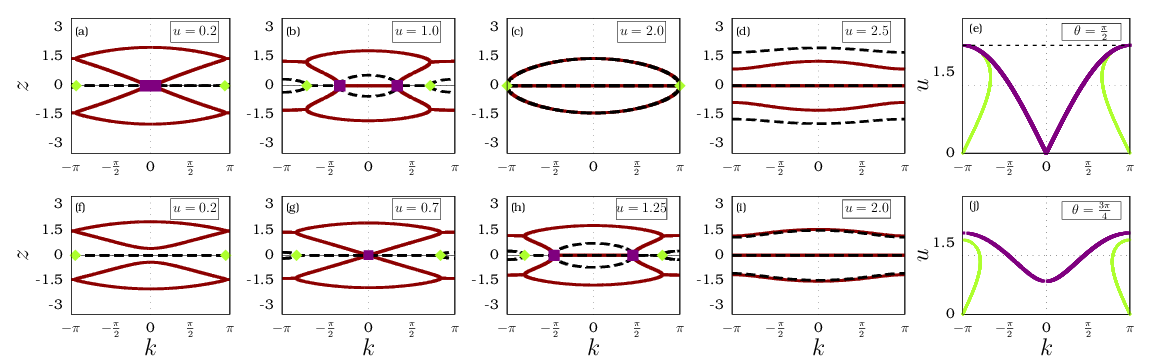}
		\caption{(a) - (d) show the band structure for $\theta = \frac{\pi}{2}$ case and (e) shows the trajectory of exceptional points EP$_1$ (shown by yellowish green curve) and EP$_2$ (shown by purple curve) for $\theta=\frac{\pi}{2}$. (f) - (i) show the band structure for $\theta = \frac{3\pi}{4}$ and (j) shows the trajectory of EP$_1$ (shown by yellowish green curve) and EP$_2$ (shown by purple curve) for $\theta= \frac{3 \pi}{4}$.  }
		\label{band_22}
	\end{center}
\end{figure}

 As the SSH Hamiltonian is Hermitian, it has real eigenspectra.  But with the introduction of non-Hermitian sites, the eigenvalues of Eq. (\ref{H_matrix}) are no longer real and are given by,
\begin{align}
	z_{1,2} =  \frac{1}{\sqrt{2}} \sqrt{X \pm \sqrt{X^2 - 4 Y}}, 
	z_{3,4} = - \frac{1}{\sqrt{2}} \sqrt{X \pm \sqrt{X^2 - 4 Y}},
	\label{eq_eig_value}
\end{align}
where
$X=2(w_1^2 + w_2^2)-u^2$ and $Y= \left(w_1^2-w_2^2 e^{-ik}\right) \left(w_1^2-w_2^2 e^{ik}\right) -w_1^2 u^2.$   The changes in the band structures with the introduced non-Hermiticity are figured out in the above mentioned three situations,  case (i): $-\frac{\pi}{2} \leq \theta \leq \frac{\pi}{2}$ (the situation in which the Hermitian SSH behaves as non-trivial insulator), case (ii): $\theta=\frac{\pi}{2}$ (the phase transition point of Hermitian SSH) and case (iii): $\frac{\pi}{2} \leq |\theta|< \pi$ (the region of trivial insulating phase in Hermitian SSH) and are presented in Figs. \ref{band_11} and \ref{band_22}. 

\par  {\bf Case (i):} Figs. \ref{band_11}(a)-\ref{band_11}(e) show the band structures for different levels of $u$ corresponding to the case $\theta=\frac{\pi}{4}$.  From Fig. \ref{band_11}(a) and from Eq. (\ref{eq_eig_value}), it is clear that the spectra becomes complex as soon as $u$ is introduced, the two yellowish green diamonds near the two extremes of Brillouin zone (BZ) represent the exceptional points ($\mathrm{EP}_1$s) beyond which the eigenvalues are complex.   With the increase of $u$, the two exceptional points $\mathrm{EP}_1$s move towards the center of BZ, where the location of these EP$_1$s can be given by 
\begin{equation}
	k=\mathrm{cos}^{-1} \left(\frac{4 u^2 w_2^2- u^4 - 8 w_1^2 w_2^2}{8 w_1^2 w_2^2}\right).
\end{equation}
Thus, the region of complex energies widened in $k$-space.  Secondly, with the increase of $u$, the band gap also get reduced.  At a critical point,
\begin{equation}
	u_{1}^c= \frac{|w_1^2-w_2^2|}{w_1},
\end{equation}
 the bands are gapless and is evident from Fig. \ref{band_11}(b).   In the case of topological systems, gap closing is often an indicator of phase transition, where the transition from trivial to non-trivial insulating phase or vice versa is expected \cite{sym}.  But, in our case, with the further increase of $u$, we find that the gap is still closed but about non-zero values of $k$ indicating the semi-metallic phase.  This arises because of the complex nature of the eigenvalues $z_3$ and $z_4$ around $k=0$, thus the degenerate point (DP) which is at $k=0$ in Fig. \ref{band_11}(b) bifurcates into two exceptional points, namely, EP$_2$s which are at
 \begin{equation}
 	k= \mathrm{cos}^{-1} \left(\frac{w_1^4+w_2^4-w_1^2 u^2}{2 w_1^2 w_2^2}\right).
 \end{equation}
  Thus, band merging occurs at these exceptional points (EP$_2$).  From Fig. \ref{band_11}(c), we note that the completely real energies exist only between EP$_1$ and EP$_2$. As $u$ increases, the EP$_2$ move towards edge of the first BZ while EP$_1$ move towards $k=0$ and as a result, at 
  \begin{equation}
  	u_{m}=\sqrt{2 (w_1^2+w_2^2)},
  \end{equation}
  EP$_1$ and EP$_2$ coalesce.  So, for $u>u_m$ (for the considered case $u_m=2.0445$), EP$_2$ only exists as shown in Fig. \ref{band_11}(d) and there is no region in $k$-space where the eigenvalues are completely real. As $u$ increases, this EP$_2$ moves towards the edge of the first BZ and it reaches the edge when,
  \begin{equation}
  	u_2^c=\frac{w_1^2+w_2^2}{w_1}.
  \end{equation}
   At this point, the bands are closed at $k=\pm \pi$.  Beyond $u=u_2^c$, there will not be any exceptional or degenerate points, the bands are gaped indicating the insulating phase (see Fig. \ref{band_11}(e)).    For instance, the location of EP$_1$ and EP$_2$ in $k-$space for different values of $u$ are given in Fig. \ref{band_11}(f), respectively using (thin) yellowish green and (thick) purple curves.   As mentioned in the above discussion, EP$_1$ alone exists for $u< u_1^c$ and at $u= u_1^c$, EP$_2$ appears.  For $u>u_1^c$, Fig. \ref{band_11}(f) indicates that the EP$_1$ and EP$_2$ moves towards each other, the region between EP$_1$ and EP$_2$, (that is, the region in which complete real energy exists) get reduced with the increase of non-Hermiticity, $u$. At $u=u_m$, EP$_1$ and EP$_2$ merge with each other and for $u_m <u <u_2^c$, EP$_2$ alone exists. Summarily, considering the case $\frac{-\pi}{2} \leq \theta \leq \frac{\pi}{2}$, we have insulating phase in the region, $0 < u \leq u_1^c$, the semi metallic phase exists in the region, 
   \begin{equation}
   	u_1^c \leq u \leq u_2^c
   \end{equation}
   and the insulating phase appears again in the region $ u > u_2^c$.

\par  {\bf Case (ii):} Now considering the case of $\theta=\frac{\pi}{2}$ ($w_1=w_2$),  the four-band spectrum of $H_k$ given in Eq. (\ref{H_matrix}) can be taken as the spectrum folded at $k= \pm \frac{\pi}{2}$ so that its band gap closes at $k=0$ \cite{antiPT}.  Increasing the value of $u$, we find band merging at non-zero values of $k$ (through bifurcation of degenerate point into exceptional points EP$_2$ denoted by purple squares in Fig. \ref{band_22}(a)) indicating semi-metallic phase.  Besides the exceptional points around $k=0$, EP$_1$ also emerges near $k=\pm \pi$ (as similar to the previous case).   EP$_1$ and EP$_2$ move towards each other as depicted by Fig. \ref{band_22}(b) and Fig. \ref{band_22}(e).  From the latter figure, we can note that upto a particular value, EP$_1$ moves towards $k=0$ but after that, it proceed towards $k= \pm \pi$.   EP$_1$ and EP$_2$ coalesce with each other when they reach $k= \pm \pi$ and this merging can be observed for $u=u_m= \sqrt{2 (w_1^2+w_2^2)}=2 w$. Fig. \ref{band_22}(c) is plotted at this point and it indicates the end of the semi metal phase.  From the Fig. \ref{band_22}(e), we can also note that this point $u=u_m$ matches with the critical point up to which the semi-metallic phase exists (ie., $u_2^c=u_m=2w$).   Thus, for $u>u_m$, the insulating phase appears where two completely real energy bands and two completely imaginary energy bands do exist as shown in Fig. \ref{band_22}(d).

\par {\bf Case (iii):} Considering the case $\frac{\pi}{2} \leq |\theta| \leq \pi$, a similar transition between insulating phases through semi-metallic phase is observed, where the insulating phase persists upto $u < u_1^c$, semi-metallic phase for $u_1^c < u < u_2^c$ and insulating phase for $u > u_2^c$.  The band structure corresponding to the different phases are presented, respectively in Figs. \ref{band_22}(f) - \ref{band_22}(i).  From these, we can observe the existence of EP$_1$ and EP$_2$ points as similar to the previous cases and the band merging in the semi-metallic phase occurs at EP$_2$.  The locations of EP$_1$ and EP$_2$ for different values of $u$ are presented in Fig. \ref{band_22}(j) which indicates that in contrast to previous cases, EP$_1$ and EP$_2$ coalesce nowhere.

\subsection{\label{zak_sec} Topological classification}
\par Now to classify the different topological phases, we calculate the Zak phase corresponding to different situations.  As shown in the literature, the Zak phase can be computed from \cite{nH_tetra_SSH, antiPT, prl_takata}, 
\begin{equation}
	\Omega = \sum_{n} i \oint \bra{\phi_n}\partial_k \ket{ \psi_n} dk,
	\label{zak_for}
\end{equation}
where the integration is taken over the complete Brillouin Zone.  Here, $n$ is the band index,  $\ket{\psi_n}$ ($\bra{\phi_n}$) is the right (left) eigenstate of the Hamiltonian $H_k$ ($H_k^\dagger$).   We wish to note that the winding number can also be calculated by averaging the Zak phase over the Brillouin zone, $W= \Omega/{2 \pi}$ \cite{prl_takata}.  
\par The unnormalized eigenfunction of $H_k$ can be given as

	\begin{equation}
		\ket{\psi_{n}} = \left(\frac{z_{n}^2 - w_1^2 + w_2^2 e^{-i k}}{w_2(z_{n}(1+e^{ik})-i u)},\frac{z_{n}^3 - iu z_{n}^2 - (w_1^2 + w_2^2)z_{n} + iu w_1^2}{w_1 w_2(z_{n}(1+e^{ik})-i u)},\frac{z_{n}(z_{n} -i u) + w_1^2 e^{ik} - w_2^2}{w_1(z_{n}(1+e^{ik})-i u)},1\right). 
		\label{righteig}
	\end{equation}

\begin{figure}[ht!]
	\captionsetup{justification=justified}
	\centering
	\includegraphics[width=1.0\linewidth]{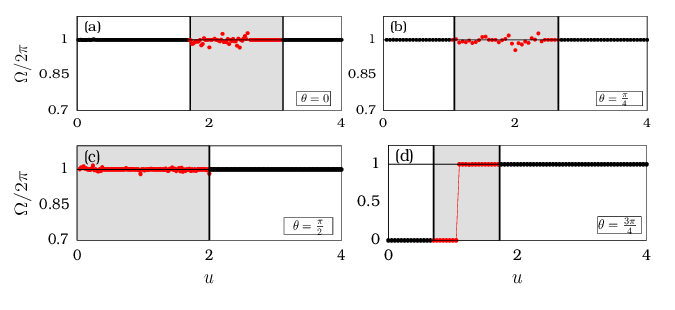}
	\caption{Zak phase values for different cases. (a) and (b) are plotted for case (i), that is, for $\theta=0$ and $\frac{\pi}{4}$. (c) and (d) are plotted for the cases $\theta=\frac{\pi}{2}$ and $\frac{3 \pi}{4}.$  }
	\label{zak_fig}
\end{figure}

One can observe that the left eigenstate of $H_k^\dagger$, $\bra{\phi_n(k)}=\ket{\psi_n(-k)}$.  Due to non-Hermiticity, the above eigenstates can be normalized through biorthonormalization ($ \bra {\phi_m(k)} \ket{\psi_n(k)}=\delta_{nm}$).   Using this biorthonormal basis, the Zak phase (vide Eq. (\ref{zak_for})) has been evaluated for different values of $\theta$ and $u$ and the results are figured out in Fig. \ref{zak_fig}. \\

\begin{figure}[ht!]
	\captionsetup{justification=justified}
	\centering
	\includegraphics[width=0.6\linewidth]{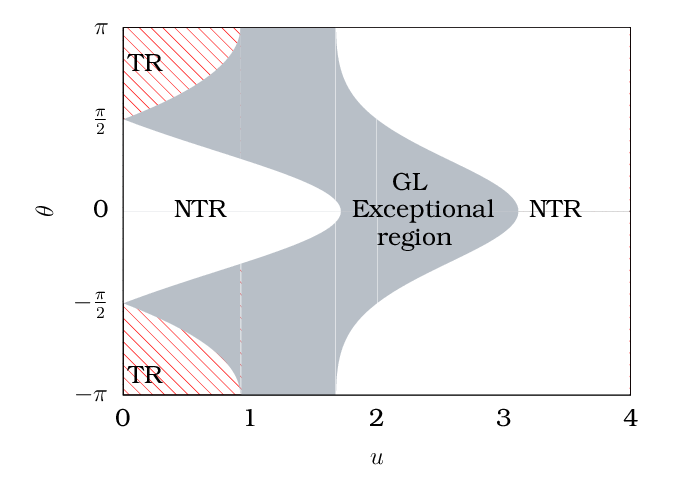}
	\caption{ Phase diagram of Eq. (\ref{syst_eq}): TR represents trivial insulating phase region, NTR represents non-trivial insulating phase region and  GL exceptional region corresponds to the gapless semi-metallic phase (which arises due to the band degeneracy at exceptional points).  Note that the eigenspectra is complex for all values of $u \neq 0$.}
	\label{phase_diag}
\end{figure}
{\bf Case (i):}  Figs. \ref{zak_fig}(a) and \ref{zak_fig}(b) are plotted respectively for $\theta=0$ and $\frac{\pi}{4}$ (corresponding to case (i)).   We recall here that in the Hermitian limit of this case, the system takes up non-trivial values for Zak phase indicating non-trivial topology. With the introduction of non-Hermiticity, for $u< u_1^c$, there is no change in the value of $\Omega$, indicating that the system remains in the non-trivial phase.  But increasing $u$ further, the previous section shows that the system will show transition to insulating phase through a region supporting gapless (semi-metallic) phase.  Thus, we expect that the intermediate gapless phase induce transition from non-trivial insultating phase to trivial insulating phase with the increase of $u$.  But in contrast to the above expectation, we observe  in Figs. \ref{zak_fig}(a) and \ref{zak_fig}(b) that the increase in $u$ induces only restoration of non-trivial insulating phase.  We can also note that the Zak phase does not approach a trivial value in the gapless phase, rather it fluctuates around $\Omega=2 \pi$ ($\frac{\Omega}{2 \pi}=1$) which indicates the topologically non-trivial nature of gapless phase.  Beyond this region, that is, in the insulating phase, $\Omega$ invariably remains at $2 \pi$. \\
{\bf Case (ii):}  Considering the particular case $\theta=\frac{\pi}{2}$ (the phase transition point in the Hermitian case), the Fig. \ref{zak_fig}(c) shows the quick increase in the values of $\Omega$ to $2 \pi$ with the introduction of $u$.  The value of $\Omega$ fluctuates around $2 \pi$ in the zero bandgap region and is invariably $2 \pi$ in the insulting region. \\
{\bf Case (iii):} Considering the case of $\frac{\pi}{2} < |\theta| \leq \pi$, Fig. \ref{zak_fig}(d) indicates that the topological trivial nature of the Hermitian case is maintained not only in the insulating phase regime ($u<u_1^c$) but also for certain values of $u$ in gapless region.  After that, a quick transition to non-trivial phase is observed.  Thus, in the region of semi-metal phase, the ones appear for smaller values of $u$ correspond to trivial phase but the ones appear for higher value of $u$ correspond to non-trivial phase. For the strong non-Hermiticity, as shown in Fig. \ref{band_22}(i), insulating phase appears for $u>u_2^c$ and the Zak phase results in Fig. \ref{zak_fig}(d) denotes that it is a non-trivial phase.  The results are summarized in Fig. \ref{phase_diag}. From the figure, we can find that in the case (i), the semi-metal phase intermediates transition between non-trivial insulating phases whereas in this case (iii), it intermediates transition from trivial to non-trivial insulating phases.  In Fig. \ref{phase_diag}, the regions of non-trivial insulating phase (NTR), trivial insulating phase (TR) and gapless (GL exceptional region) regions are figured out in the $\theta - u$ parametric space.
\par With this, let us see the nature of topological states in the situation with open boundary condition and check whether the non-trivial phases observed in different parametric region support robust edge modes.

\section{\label{sec_open}Open boundary condition}

\begin{figure*}[htb!]
	\captionsetup{justification=justified}
	\begin{center}
		\includegraphics[width=1.0\linewidth]{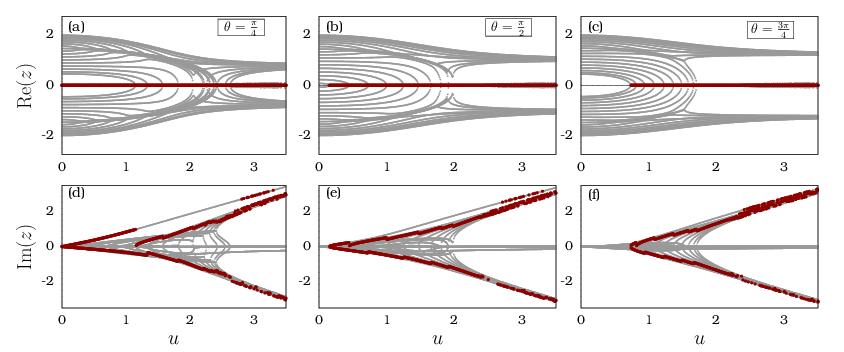}
		\caption{Real and Imaginary parts of $z$ vs $u$ in the open boundary conditions. (a) and (d) are plotted for the case $\theta = \frac{\pi}{4}$. (b) and (e) for $\theta=\frac{\pi}{2}$ and (c) and (f) for $\theta=\frac{3 \pi}{4}$. Here, the dark red dots are used to distinguish the zero edge mode from the other eigenvalues.}
		\label{zvsuu}
	\end{center}
\end{figure*}
\par Considering the open boundary conditions, we study the existence of edge modes and the possibility of non-Hermitian skin effect \cite{long-rangeSSH, Xu, KZhang, Hou}.  For instance, the spectra related to the three different cases are given in Figs. \ref{zvsuu}(a)-\ref{zvsuu}(f).  Due to the chiral symmetric nature of the system, we can observe that the eigenspectra is symmetric with respect to $z=0$ axis.  First considering the case of $\theta= \frac{\pi}{4}$ (Fig. \ref{zvsuu}(a)), we find that the bands are gaped for smaller values of $u$, then the gapless modes exist for some intermediate values of $u$ and the band gap is established again for higher values of $u$.  Considering the case of $\theta=\frac{\pi}{2}$ (Figs. \ref{zvsuu}(b)), we can observe transition from gapless phase to insulating phase with the increase of $u$.  For the case $\theta=\frac{3 \pi}{4}$, Fig. \ref{zvsuu}(c) shows that the gapless phase appears for some intermediate value of $u$ and for other values of $u$, the bands are gaped. Now, focusing on the zero eigenmode, Figs. \ref{zvsuu}(a) shows that in the case of $\theta= \frac{\pi}{4}$, this mode exists for all values of $u$. But in the case of $\theta= 3 \pi/4$ (from Fig. 6(c)), zero mode does not exist for smaller values of $u$ as it corresponds to trivial region. 
\begin{figure}[ht!]
	\captionsetup{justification=justified}
	\begin{center}
		\includegraphics[width=1.0\linewidth]{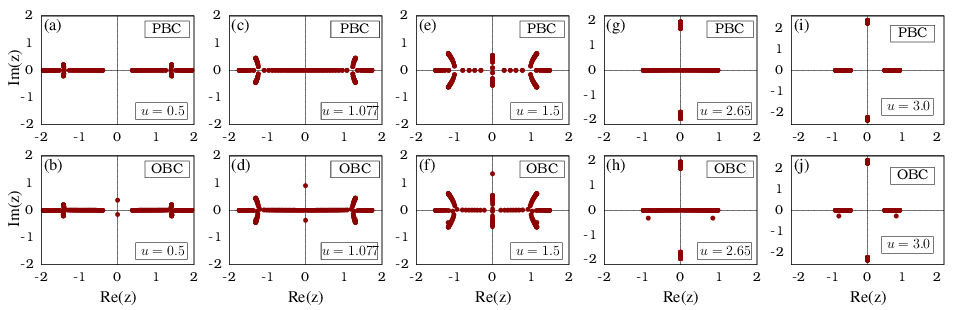}
		\caption{Comparison of the band structures in the cases of OBC and PBC for $\theta=\frac{\pi}{4}$.  }
		\label{pobc1}
	\end{center}
\end{figure}

\begin{figure}[ht!]
	\captionsetup{justification=justified}
	\begin{center}
		\includegraphics[width=1.0\linewidth]{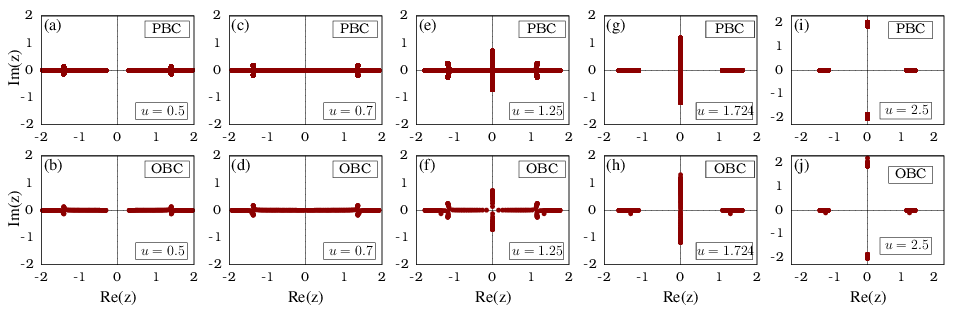}
		\caption{Comparison of the band structures in the cases of OBC and PBC for $\theta=\frac{\pi}{4}$. }
		\label{pobc2}
	\end{center}
\end{figure}
  The above observations show that they are qualitatively on par with the results of PBC. But, to have more detailed study on the sensitivity of system towards boundary conditions, we plot the band structures in OBC and PBC situations in Fig. \ref{pobc1} (for $\theta= \frac{\pi}{4}$) and Fig. \ref{pobc2} (for $\theta = \frac{3\pi}{4}$). From these figures, it is obvious that the system has line gap in all the insulating phase regimes and it indicates that the system can be transformed adiabatically to a Hermitian system.   From Fig. \ref{pobc1}(a) and \ref{pobc1}(b) (which is plotted $u=0.5$, that is, in the NTR region for lower values of $u$ in Fig. \ref{phase_diag}), we can find that the band structures corresponding to OBC and PBC looks same and the zero edge modes present at the forbidden band gap of PBC. The figure shows that there is no violation in bulk-boundary correspondence.  However, considering the cases $u=1.077$ and $u=1.5$ (from Figs. (\ref{pobc1}(c)-\ref{pobc1}(f)), around the GL region/non-trivial semi-metallic region, the bulk-boundary correspondence is weakly preserved, where we can find the variations in the band structures corresponding to OBC and PBC. We can also note the emergence of non-zero discrete modes in the figure.   But considering Fig. \ref{pobc1}(g) -  \ref{pobc1}(j)  corresponding to $u=2.65$ and $u=3.0$, the similarities in the band structures indicate the preservation of bulk-boundary correspondence. Similar behavior is observed in the case of $\theta=\frac{3 \pi}{4}$, where the edge modes are absent for lower values of $u$ due to the trivial nature of the system (vide Figs. \ref{pobc2}(a)-\ref{pobc2}(b)). Difference in the band structures is observed for the values of $u$ near the semi-metallic region (vide Figs. \ref{pobc2}(c)-\ref{pobc2}(f)).  Similarity in the eigenspectra can be seen for higher values of $u$ (from Figs. \ref{pobc2}(g)-\ref{pobc2}(j)).  Thus, even in this case, the violation of bulk-boundary correspondence can be seen at the boundaries of semi-metallic region. 
\par
\begin{figure}[ht!]
	\captionsetup{justification=justified}
	\begin{center}
		\includegraphics[width=1.0\linewidth]{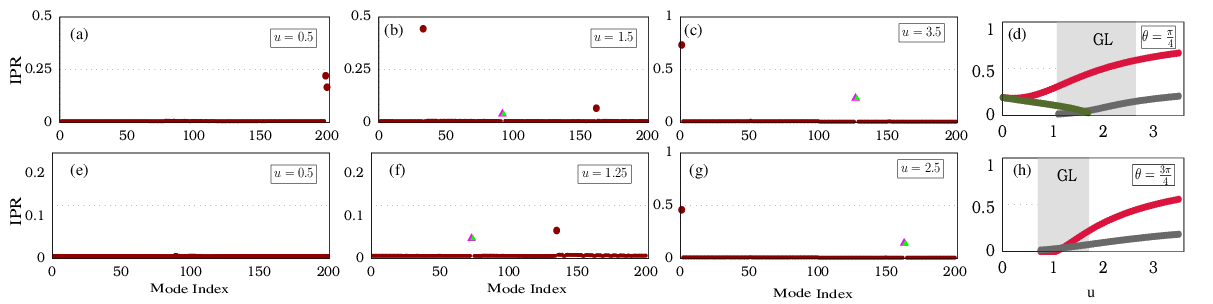}
		\caption{ (a) - (d) Inverse participation ratio for different bulk and edge modes for the case $\theta=\pi/4$ and (e)-(h) for $\theta=3 \pi/4$.  The triangles in (a)-(c) and (e)-(g) represent non-zero right edge mode and the isolated dark-red dots represent the left and right zero edge modes. The IPR values for different $u$ corresponding to left zero edge mode (red curve), right zero edge mode (dark-green curve) and right non-zero edge modes (gray curve) are given in (d) and (h).  }
		\label{ipr}
	\end{center}
\end{figure}

 The next question along this line is the possibility of non-Hermitian skin effect where macroscopic number of modes localize at one of the boundaries of the lattice, showcasing the non-reciprocal nature of non-Hermitian systems.  For $\theta= \frac{\pi}{4}$, the system support left and right edge modes for $u=0$.  The inverse participation ratio (IPR) (which measures the number of sites in which the edge mode is localized) are obtained for different bulk and boundary modes {\cite{IPR, Kvande}}. For $u=0.5$ and $\theta= \pi/4$, the Fig. \ref{ipr}(a) shows that the value of IPR is high for two states which are zero energy edge modes where one of the modes represents left edge mode and the other represents right edge mode. Increasing the value of $u$ (from Fig. \ref{ipr}(a)), the IPR value of right edge mode decreases and this zero right edge mode delocalizes at the edges of GL exceptional region. Beyond the boundary of GL region, non-zero edge modes arises with moderate values of IPR.  These non-zero edge modes (which are complex conjugate to each other) are found to localize at the right boundaries of the lattice.  Their IPR value increases as $u$ increases (as shown in Fig. \ref{ipr}(c)). Fig. \ref{ipr}(d) shows the variation in the IPR values of zero and non-zero edge modes where one can find the delocalization of right edge mode to the bulk near the boundaries of GL region and the emergence of two non-zero right edge modes beyond the boundaries of GL region. Thus both left edge mode and right edge modes exist in almost all the region except near the boundary of GL region (where the violation in bulk-boundary correspondence is observed).  In all the regions, left edge mode is highly localized compared to the right edge mode.  Even though we could not observe localization of macroscopic number of edge modes in one of the edges, the difference in the IPR values of left and right edge mode indicates a sort of non-reciprocal nature which can be associated with the non-Hermitian skin effect. We recall here that in the non-Hermitian SSH lattice with all sites to be non-Hermitian \cite{indian}, the left and right edge modes have same IPR values and exhibit no non-Hermitian skin effect. 
  A similar weak non-Hermitian skin effect is observed in the case of $\theta=3 \pi/4$, where edge modes are absent in the trivial insulating regime as shown by Fig. \ref{ipr}(e).  Fig. \ref{ipr}(f)-\ref{ipr}(h) represents the emergence of zero left edge mode and non-zero edge mode (corresponding to right edge localization) and their IPR values increases with respect to $u$.  In this case, the zero right edge mode does not arise in any of the parametric region.   
\par We have also tested the robustness of these observed edge modes by adding perturbation to the Hamiltonian in the form,
	\begin{equation}
	H_p(k)= \begin{pmatrix}
		...  & . & . & . & . & . & ...\\
		...  & . & . & . & . & . & ...\\
		...  & w_2 \epsilon_{w,4l-4} & 0 & 0 & 0 & 0 & ...\\
		...  & i u \epsilon_{u,4l-3} & w_1 \epsilon_{w,4l-3} & 0 & 0 & 0 & ...\\
		...  & w_1 \epsilon_{w,4l-3} & - i u \epsilon_{u,4l-2} & w_2 \epsilon_{w,4l-2} & 0 & 0 & ...\\
		...  & 0 & w_2 \epsilon_{w,4l-2} & 0 & w_1 \epsilon_{w,4l-1} & 0 & ...\\
		...  & 0 & 0 & w_1 \epsilon_{w,4l-1} & 0 & w_2 \epsilon_{w,4l} & ...\\
		...  & . & . & . & . & . & ...\\
		...  & . & . & . & . & . & ...\\
	\end{pmatrix},
\end{equation}
where $l$ is the cell index and the fluctuation coefficients $\{\epsilon_{w, j}\}$ or $\{\epsilon_{u, j}\}$, ($j=4l-x$ is the site index) are considered to take random values which are distributed according to Gaussian distribution with its mean equal to $1$ and standard deviation $\sigma = 0.2$.  The eigenstates of this Hamiltonian are found in order to study the robustness of the edge modes.
 \begin{figure}[ht!]
	\captionsetup{justification=justified}
	\begin{center}
		\includegraphics[width=1.0\linewidth]{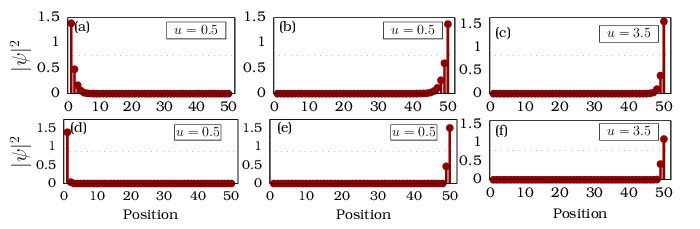}
		\caption{Edge states observed in the lattice for open boundary conditions in the case of $\theta= \frac{\pi}{4}$. (a) and (b) The left and right zero edge modes observed for $u=0.5$ and right non zero edge mode for $u=3.5$. (d)-(f) shows the Robustness edge modes. }
		\label{edgy}
	\end{center}
\end{figure}
\par For example, the results corresponding to the edge mode of $\theta=\frac{\pi}{4}$ case is presented in Fig. \ref{edgy}. Fig. \ref{edgy}(a), \ref{edgy}(b) and \ref{edgy}(c) respectively shows the left zero edge mode, right zero edge mode and right non-zero edge mode. The corresponding edge modes are plotted from the eigenstates of perturbed Hamiltonian respectively in Fig. \ref{edgy}(d), \ref{edgy}(e) and \ref{edgy}(f).  The figure shows that the edge modes do exist even under perturbation which shows the robust nature of the observed edge modes.  We also note here that the edge states observed in the gapless region are also found to be stable.

\section{Summary}
\par In this article, we studied a non-Hermitian variant of SSH model in which the strong non-Hermiticity favors non-trivial insulating phase.  We observe interesting form of phase transition which is mediated by semi-metallic phase where the system shows two different phase transitions.  In one situation, the system transits from trivial insulating phase to non-trivial insulating phase through semi-metallic phase (which is found to be topologically trivial for smaller values of $u$ and non-trivial for higher values of $u$).  Interesting phase transition is observed in the other situation, where the system in the non-trivial insulating phase (in the absence of $u$) shows transition to non-trivial semi-metal phase.  In this case, increase in the non-Hermiticity again restores non-trivial insulating phase.  
\par Considering the lattice with open boundary condition, we study the existence of robust edge modes in different non-trivial regions including the gapless region. Considering the case in which unusual transition between non-trivial insulating phases occurs, we looked for the possible difference that could be observed in the topological states corresponding to the two non-trivial insulating phases (which appears in the two extremes of semi-metal region) observed in the two extreme values of non-Hermiticity. The non-trivial phase which appears for smaller values of $u$ support two fold degenerate zero energy edge modes, namely, the left and right edge modes.  However, the ones observed for high non-Hermiticity are left zero edge mode and right non-zero edge modes.  The edge modes observed in different parametric regions (including the ones observed in the non-trivial gapless phase) are found to be robust against perturbations. We have also illustrated that our system exhibits weak form of non-Hermitian skin effect where the left edge modes are more localized compared to right edge modes. This is in contrast to the SSH model with completely non-Hermitian sites shown in \cite{indian}, where both left and right edge modes have same IPR values and exhibit no non-Hermitian skin effect.  By considering staggered Hermitian and non-Hermitian dimers, we can see a weak form of non-Hermitian skin effect.   The studies can be extended to study lattices with lesser number of non-Hermitian sites.  As the mentioned Hamiltonian can be observed in the photonic lattice situation, the experimental studies on the system may reveal interesting features and applications.
  
\section{Acknowledgement}

AN wishes to thank MoE-RUSA 2.0 Physical Sciences, Government of India for providing a fellowship to carry out this work. The work of MS was supported by DST-SERB, Government of India, under the Grant No. CRG/2021/002428 and MS also acknowledges Council of Scientific and Industrial Research (CSIR) for research project sponsored under the Grant No. 03/1482/2023/EMR-II.

\end{document}